# Depletion of Gaseous CO in Protoplanetary Disks by Surface Energy Regulated Ice Formation

**Authors:** Diana Powell[1,2,*], Peter Gao[3], Ruth Murray-Clay[4], Xi Zhang[5]

**Affiliations:**

[1]Harvard & Smithsonian | Center for Astrophysics, Cambridge, MA 02138, USA

[2]NHFP Sagan Fellow

[3]Earth and Planets Laboratory, Carnegie Institution for Science, Washington, DC 20015, USA

[4]University of California Santa Cruz, Department of Astronomy and Astrophysics, Santa Cruz, CA 95064, USA

[5]University of California Santa Cruz, Department of Earth and Planetary Sciences, Santa Cruz, 95064, USA

*Correspondence to: diana.powell@cfa.harvard.edu

Empirical constraints of fundamental properties of protoplanetary disks are essential for understanding planet formation and planetary properties[1,2]. Carbon monoxide (CO) gas is often used to constrain disk properties[3]. However, estimates show that the CO gas abundance in disks is depleted relative to expected values[4-7], and models of various disk processes impacting the CO abundance could not explain this depletion on observed ~1 Myr timescales[8-14]. Here we demonstrate that surface energy effects on particles in disks, such as the Kelvin effect, that arise when ice heterogeneously nucleates onto an existing particle, can efficiently trap CO in its ice phase. In previous ice formation models, CO gas was released when small ice-coated particle where lofted to warmed disk layers. Our model can reproduce the observed abundance, distribution, and time evolution of gaseous CO in the four most studied protoplanetary disks[7]. We constrain the solid and gaseous CO inventory at the midplane and disk diffusivities and resolve inconsistencies in estimates of the disk mass—three crucial parameters that control planetary formation.

With a consideration of surface energy effects such as the Kelvin effect (see methods), the process of CO ice nucleation in protoplanetary disks requires significantly larger supersaturations of CO gas ($S > \sim 100$) than the process of condensational growth ($S > \sim 1$). Thus, once nucleation occurs, condensational growth onto the nucleated particles quickly becomes the dominant ice formation process and rapidly depletes the CO gas in the regions of the disk near the midplane where it is initially supersaturated, creating a vertical concentration gradient (Fig. 1a). This concentration gradient is reduced by diffusive mixing which funnels CO gas from the upper warm layers to the disk midplane. The ongoing process of ice condensation readily depletes the fresh supply of CO gas such that it does not reach the critical supersaturation needed to nucleate new ice particles throughout the bulk of the disk after the first few thousand years of the disk's lifetime. The initial fraction of particles that are coated in CO ice is thus primarily set at this early stage. CO gas is incorporated into solid CO on a timescale that is relatively fast once gas diffuses towards the disk midplane, thus the CO gas mole fraction will be significantly depleted throughout the vertical extent of the disk (Fig. 1a) over a vertical diffusion timescale (see Supplementary Figure 3).

The progressive depletion of CO gas is facilitated by the condensation of CO onto large CO-coated ice particles which have the lowest surface energy barriers (see methods). These large ice particles form from the initial population of CO-nucleated ice particles that grow, first via condensational growth and later via coagulation, before eventually drifting radially inwards towards the host star (see methods). For evolved disks, most of the solid CO mass is therefore located on the largest ice-coated particles in the size distribution (Fig. 1b). These large ice particles settle below the warm layers to a region of the disk that is too cold for CO evaporation and is optically thick to UV photons, such that ice photodesorption does not occur (Fig. 1b). This settling of large grains is prevalent throughout the ice forming regions of the disk (Fig. 2a), leading to a large sequestration of volatile material in the disk midplane.

If the effects of surface energy on ice nucleation are neglected, ice preferentially and continuously forms on small grains that are lofted efficiently to warm layers, where they release CO ice back into the gas phase (see Supplementary Figure 2) and limit the maximum amount of gaseous CO depletion (Fig. 1b). In other words, by considering surface energy effects that largely prevent small grains from bearing CO ice, we circumvent the problem of CO resupply to the warm layers through lofting and evaporation of small CO ice grains (Fig. 1b).

The amount of observed CO gas in protoplanetary disks varies with semi-major axis and the disk age. Depletion occurs quickly immediately exterior to the midplane CO ice line because gas can quickly diffuse to the disk midplane and form ice (Fig. 2b). The vertical diffusion timescale increases in the outer disk where the scale heights are large due to disk flaring such that the timescale of depletion varies by orders of magnitude across a given system (see Supplementary Figure 3). As a result, while the depletion of CO at the midplane in the outer disk is large due to the low midplane temperature, the observed depletion in the upper warmer layer will be minimal if the vertical diffusion timescale is longer than the current age of the system. Thus, in the outer disk, the amount of observed depletion varies with semi-major axis and increases with time such that the disk will become more depleted in gaseous CO as it evolves (Fig. 2b). In other words, the amount of CO gas depletion in the outer disk is simply a function of the fraction of the disk diffusion timescale that has elapsed since the system's birth (Fig. 3). This is in agreement with recent observations of gaseous CO where very young disks (<1 Myr) do not appear to be depleted in CO gas while older systems (>1 Myr) can be depleted significantly[8,9]. The observed timescale of depletion can be naturally explained by this model, while previous ice formation models and chemical models with ISM cosmic-ray rates generally require ~3 Myr to deplete CO by an order of magnitude given expected cosmic ray abundances[11,13,14]. Depending on the speed of particle growth and drift there can also be a local enhancement of CO gas around the ice line once solid CO ice particles have drifted inwards and released their volatile material. This feature is initially very narrow in radial extent and is diffusively broadened with time (Fig. 2b). The CO-enriched gas interior to the disk critical semi-major axis is then accreted onto the host star.

We model four systems and compare to observations in Fig. 4. We constrain the bulk disk diffusive properties and the composition of solid and gaseous material in the disk midplane as a function of both semi-major axis and time. The observed depletion is not always directly correlated with disk age in the observed sample, as some young disks are very depleted in CO gas while some older disks are not[15]. The amount of depletion from initial values depends sensitively on the diffusion present in the system. Thus, in addition to the system age, the primary factor that controls the distribution of gaseous CO is the level of diffusion in the disk. We thus optimize the fit of our model to the data by varying the level of diffusion present in the system (see methods). Our empirical constraints of the diffusion parameter (Table 1) all fall within a range of plausible values[3] and are qualitatively consistent with estimates from non-thermal gas velocities[16-18] (see methods).

While the relative CO abundance depends sensitively on the diffusion present in the system, the total CO mass present in the system depends on the total disk mass. This modeling demonstrates that CO gas is indeed depleted in systems such that a simple conversion from CO gas emission to other fundamental properties such as the total disk mass is not appropriate without modeling non-equilibrium ice formation including surface energy effects. With the modeling presented in this work, disk masses estimated from CO agree with those estimated from dust line modeling[19-21], lower limits from HD observations (when relevant)[5,22], and dust emission (with a dust-to-gas ratio of ~ $10^{-3}$ appropriate when considering grain growth and particle drift[20]).

We constrain the inventory of CO in gas and solids in the protoplanetary disks considered here (Table 1). Our results are consistent with an initial CO abundance in these systems that is similar to the interstellar ratio (which we take to be $10^{-4}$ by number[23]). The partitioning of the CO mass inventory depends on the disk diffusive parameter, $t_{age}/t_{diff}$, where $t_{age}$ is the age of the system and $t_{diff}$ is the diffusion timescale (see methods). When the diffusive parameter is very small, as is the case for the disk around HD 163296, most of the initial CO mass remains in the gas phase in the outer disk where it is being converted to CO ice as described in the first step in Supplementary Figure 1. As the diffusive parameter increases, the ice formation process continues and most of the initial CO mass will be located in solid ice in the outer disk, as is the case for the disk around DM Tau. At larger diffusive parameters, the ice particles that have depleted the gaseous CO in the outer disk have drifted inwards past the CO ice line where they desorb their volatile materials (steps 2-4 in Supplementary Figure 1). Thus, disks with larger diffusive parameters have either a roughly equal amount of their initial CO mass located in gas in the inner disk and lost to accretion, like the disk around TW Hya, or have lost a majority of their initial CO mass to accretion, like the disk around IM Lup. We note that the age of protoplanetary disks is uncertain and literature estimates can vary by a few million years for a given system. In our modeling, the age of the system is degenerate with the level of diffusive mixing we derive such that an older, less-diffusive disk like HD 163296, can have similar levels of observed depletion as a significantly younger, more-diffusive disk like DM Tau, though the corresponding distribution of CO mass throughout the disk will vary between these two cases.

Our results demonstrate that the abundance of the gaseous and solid material in protoplanetary disks in the planet forming regions changes as a function of time and disk semi-major axis. For the solid material, the fraction of ice relative to refractory solids in the outer disk is a function of time and particle size. At late times probed by observations, the total mass of CO ice-free grains is roughly an order of magnitude larger than the total mass of CO ice particles due to inward drift of large CO ice grains. At early times, however, the bulk ice to refractory ratio is larger due to an ice fraction on large particles that is much higher than the bulk ice to refractory ratio. Large bodies that form in protoplanetary disks at early times via the processes of streaming instability or pebble accretion, which preferentially accrete the largest particles in the distributions presented in this work, are likely to have compositions with significant volatile components.

A prediction of this model is that the ~micron sized particles in evolved disks are not likely to have a stable coating of CO ice (Fig. 1a). The population of small CO ice-free grains that is abundant throughout the vertical and radial extent of the disk is likely to dominate the reflected light flux and possess scattering properties different from those of larger CO-rich particles. Future observational studies, using JWST[24] and other facilities, of the lack of infrared CO ice-features or CO ice-rich reflective properties contributed from small grains could validate this prediction (see methods).

A similar process of depletion should occur for the other volatile species in disks, such as $H_2O$ and $CO_2$—meaning that future modeling can constrain the abundance ratios in disks. The abundance ratios of different elements, such as the C/O ratio, can be used to determine a planet's formation region via comparing the abundances of a planet to the abundance ratios as a function of radius in protoplanetary disks. Exterior to the CO ice line, we expect that the bulk composition of the ice particles at the midplane will be a stellar abundance and similarly, the bulk composition of the ice particles between the $CO_2$ and CO ice lines should be sub-stellar in C/O as predicted in [2]. This is because our modeling indicates that the ice formation of volatile species in these regions is regulated by the same vertical diffusion timescale, even though these species have different ice formation efficiencies. Because giant planets that form via core-accretion are thought to obtain the majority of their metallicity primarily from solid material, the compositions of young, giant planets that span this region should see a corresponding change in C/O ratios. Indications of this trend are emerging for the planets in the HR8799 planetary system where the two planets in the outer solar system (b and c) may have stellar C/O ratios[25-28]. While the C/O ratios of the two inner planets are not as well constrained, there have been some indications of substellar C/O ratios[27] which agrees with these predictions. The C/O ratios should deviate from those predicted in[2] interior to the $CO_2$ ice line. Qualitatively, this will occur because the radial transport rates of solid $CO_2$ and gaseous CO will differ though further work is needed to robustly constrain abundance ratios in the inner disk.

Including surface energy effects in ice nucleation and condensation in future disk models will be important for understanding the composition of solids of different sizes, particularly at later times once particle drift is efficient. This is because species with larger surface energies will more readily form on larger particles that are more susceptible to drift which could lead to a solid abundance that varies with particle size and disk semi-major axis. Once gas has been depleted on a vertical diffusion timescale, the gas composition will be set by the saturation vapor pressures of the various gas species and should differ significantly from stellar abundance ratios though the overall gas metallicity will be greatly reduced. Our work provides new constraints on the environment of planet formation and sheds light on the variation of the bulk composition of planetary bodies and their building blocks with orbital distance and formation timescale.

## Tables

| Disk | Disk Mass [$M_\odot$] | α | Accreted CO Mass [%] | Inner CO Gas Mass [%] | Outer CO Gas Mass [%] | CO Ice Mass (outer disk) [%] |
|---|---|---|---|---|---|---|
| TW Hya | 0.11 ([20,21]) | $8 \times 10^{-4}$ | 51 | 47 | 2 | <1 |
| HD 163296 | 0.21 ([20]) | $10^{-4}$ | 4 | 16 | 62 | 18 |
| DM Tau | 0.05 ([22]) | $3 \times 10^{-3}$ | 14 | 12 | 24 | 50 |
| IM Lup | 0.10 (see methods) | $1.5 \times 10^{-2}$ | 69 | 17 | 2 | 12 |

**Table 1.** Model constraints and outputs: the total disk mass, globally averaged turbulent viscosity parameter, and the percentage of the initial CO mass that has been accreted onto the host star, currently resides in gaseous form interior to the midplane ice line, currently resides in gaseous form exterior to the midplane ice line, and is currently located in solid form exterior to the midplane ice line.

**Figure Legends/Captions**

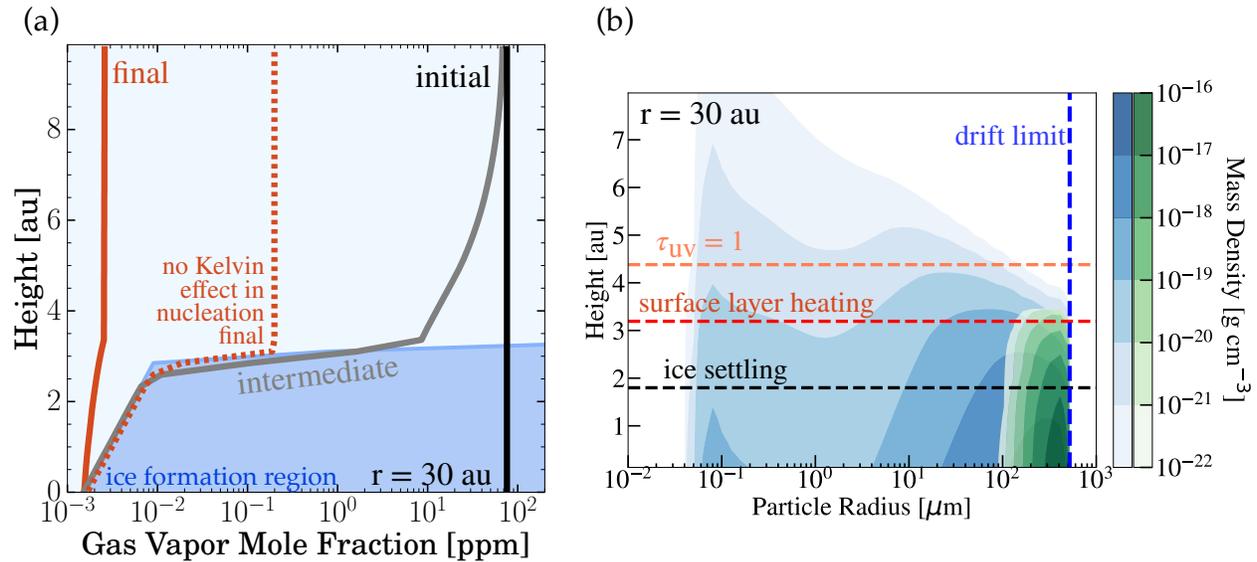

**Fig. 1. The distribution of gaseous CO is regulated by preferential condensation of CO onto large particles.** (a) The initial CO abundance (black line) is first depleted (gray line) in the regions where it is supersaturated (dark blue shaded region). CO from the upper layers is then mixed downwards and depleted on a vertical diffusion timescale. Without radial resupply, the final CO gas mole fraction will be constant with height and fixed to the midplane CO saturation mole fraction (solid coral line). If the Kelvin effect is not included, then CO would condense on small grains as well. These small grains can be lofted and will release their volatile components in the warm upper layers, thus limiting the maximum amount of depletion possible at a given semi-major axis (dashed coral line). (b) A fiducial distribution of ice coated particles (green) and ice-free grains (blue) at a given disk semi-major axis (r = 30 au, here for TW Hya at 5 Myr). The growth of the large particles is limited by particle drift (blue dashed line). The remaining CO ice-coated particles settle to the disk midplane (black dashed line) and are not readily lofted to the upper regions of the disk where surface layer heating occurs (red dashed line) or where the disk becomes optically thin to UV-photons (the UV optical depth, $\tau_{UV} = 1$, dashed orange line). The ice particles are thus unable to release their volatile material due to either evaporation or photodesorption.

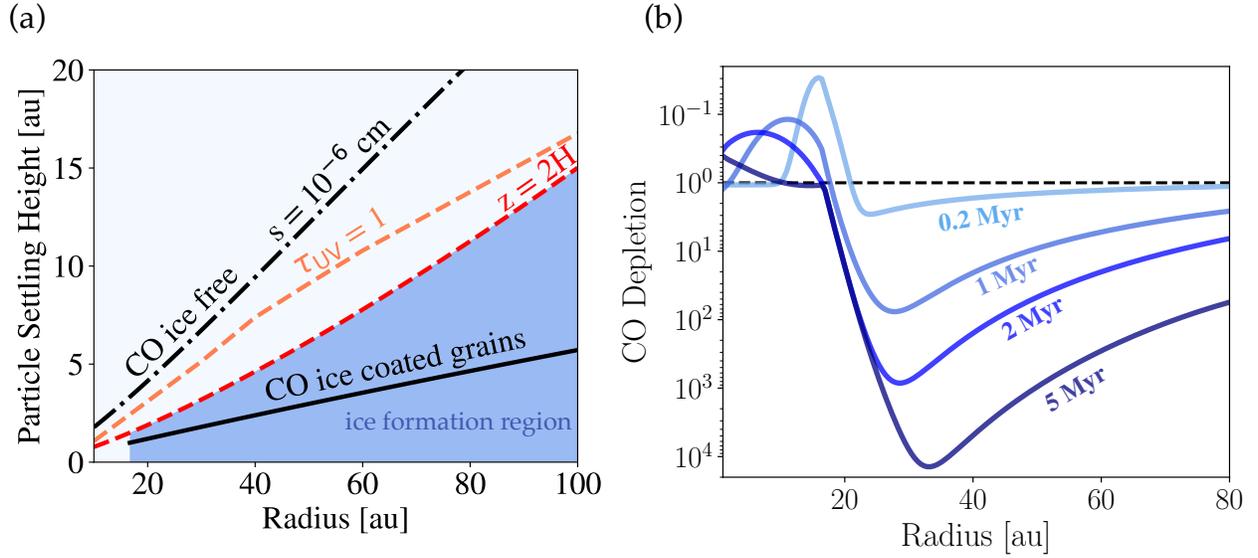

**Fig. 2. The radial evolution of CO in the disk around TW Hya:** (a) Ice-coated particles settle vertically throughout the radial extent of the disk (ice scale height is the black solid line, small particle scale height is the dash-dot line) below the surface layer heated region (i.e., the ice formation region, dark blue) which we approximate as beginning at an altitude, z, that is two scale heights, H, above the midplane (red dashed line) based on more detailed modeling and observations[29,30]. The ice-coated particles also remain below the region that is optically thin (with a UV optical depth, $\tau_{UV} < 1$) to UV photons (orange dashed line), allowing CO to be sequestered in solids. (b) The amount of CO gas depletion defined as the initial (interstellar) gas-phase CO-to-$H_2$ ratio divided by the final ratio (blue lines) in protoplanetary disks varies with semi-major axis and increases with time such that the disk will become more depleted in gaseous CO in the outer disk as it evolves. The results were calculated using disk parameters for the disk around TW Hya (see methods).

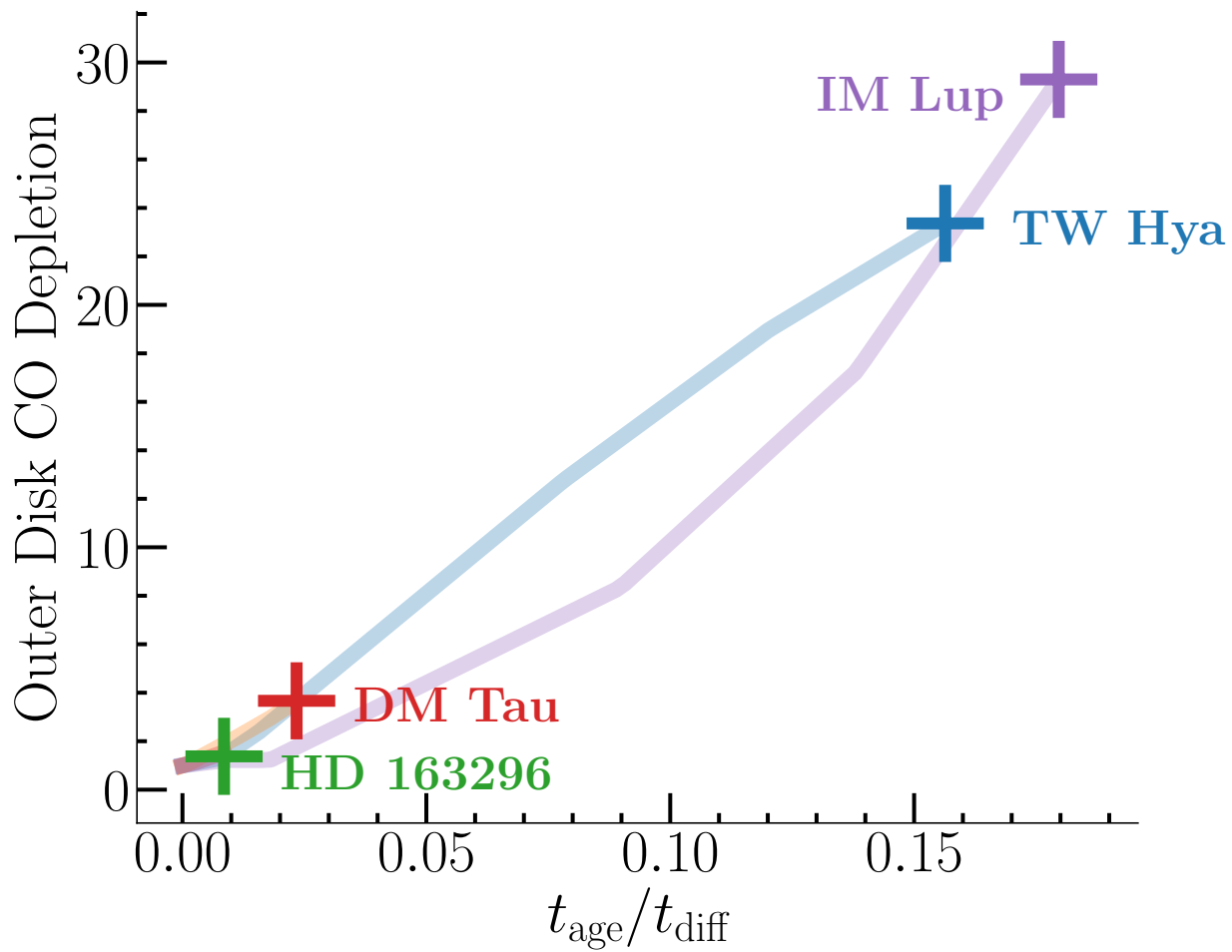

**Fig. 3. The amount of CO gas depletion depends on the disk diffusion timescale.** The average CO depletion (see Fig. 2) exterior to the midplane ice line is a function of the current disk age divided by the disk diffusion timescale. The amount of gaseous CO depletion increases with time (colored lines indicate depletion evolution).

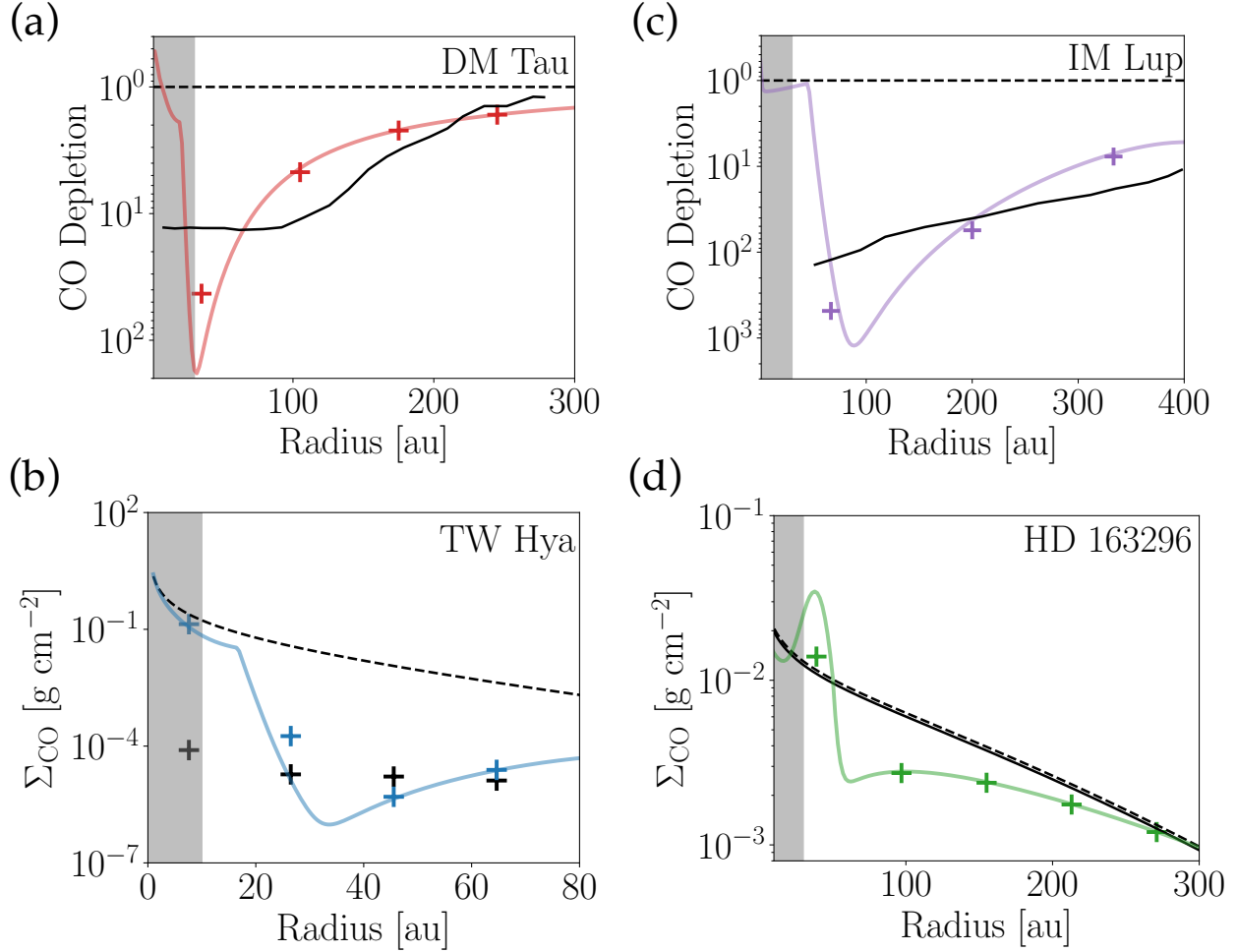

**Fig. 4. Surface-energy-regulated ice formation can describe CO gas abundances for a variety of observed disks.** For each system we compare our model results (colored lines and crosses indicating the beam-size-convolved value) to the quantity reported in the observational literature (black solid lines or crosses indicating the half-beam separation). The literature reported quantity is either CO surface density or CO depletion and we have presented our results to be consistent with the original reported quantity (see Fig. 2). The initial CO gas distribution is also shown (black dashed lines). The CO emission that arises from the gray-shaded regions is likely optically thick throughout the vertical extent of the disk for all commonly observed CO isotopologues, unless clearing occurs by gas-planet interactions, which are not considered in this model. For a detailed discussion of our model-data comparisons see methods. (a) DM Tau[6, 22], (b) IM Lup[7] (c) TW Hya[6], (d) HD 163296: disk-averaged values indicate a lack of CO depletion while higher spatial resolution data show some features[7, 15, 31, 32].

## Methods

We model the formation of CO ice in a series of vertical one-dimensional (1D) columns for a discrete set of disk semi-major axes. Each vertical disk model starts with an initial population of small grains that are allowed to grow via coagulation and the nucleation and condensation of CO ice. The model is then coupled with a global model of gaseous diffusion and radial dust grain aerodynamics. In particular, we account for accretion onto the star from the disk as well as the release of gaseous CO due to the influx of icy pebbles from the disk's outer regions into its hotter, inner regions.

The primary inputs to our modeling are the total gaseous surface density of the protoplanetary disk, the midplane temperature structure, the age of the system, and the diffusion coefficient that describes the mixing in the system. The diffusion coefficient is treated as a free parameter, as it is not significantly constrained from observations. As noted in the main text, our derived values are consistent with existing estimates[16-18] and within the range of reasonably expected values.

<u>Non-Equilibrium Ice Formation and Particle Evolution</u>

We model a series of 1D vertical columns separated by 10 au exterior to the midplane CO ice-line with an additional vertical column immediately radially exterior to the midplane ice-line. The results presented in this work do not change when the radial resolution of the models is doubled. At each vertical column we sample the disk density and midplane temperature (for values see the Disk Parameters section) at that location and use these values as inputs in our modeling of vertical particle evolution.

The primary quantities of interest from the modeling of CO ice formation are the extent to which gas can be vertically depleted in the column at a given semi-major axis, the solid ice to gas ratio, and the resulting size distribution of the particles at that location. To derive these quantities, it is essential to treat ice formation as a non-equilibrium process and to fully resolve the size distribution of particles.

We modify the 1D Community Aerosol and Radiation Model for Atmospheres (CARMA) to simulate the formation of CO ice in the diffuse outer regions of protoplanetary disks. CARMA computes the vertical and size distributions of ice particles by solving the discretized aerosol continuity equation, taking into account particle nucleation (homogeneous and heterogeneous), condensation, evaporation, and coagulation. CARMA uses bins to resolve the particle size distribution, allowing for multiple particle modes to be simulated simultaneously and avoiding the need to parameterize the size distribution using an analytical function. We refer the reader to[33] and the appendix of[34] for an additional detailed description of CARMA and its history. We do not consider particle fragmentation or the photodesorption of CO ice. For the regions of the disk where the formation of CO ice occurs (typically beyond ~10 au), the process of fragmentation is not likely to contribute to the shaping of the particle size distribution assuming reasonable levels of disk turbulence and particle compositions [35]. We do not include the photodesorption of CO ice as all of the ice-covered particles in our modeling evaporate quickly once they reach the heated surface layers, which is often several au deeper in the disk than the region that is optically thin to UV photons (see Fig. 2a). We approximate the UV optical depth using the method described in [12] (see their Equation 7).

For each column, we begin with an initial interstellar medium (ISM) abundance (dust-to-gas mass ratio of $10^{-2}$) of small, $10^{-6}$ cm, grains. This initial grain size is smaller than the ~ 0.1 μm size of a typical ISM grain though this assumption does not affect the resulting particle size distribution. These particles are allowed to coagulate and grow to larger sizes (see appendix section A.3 of [34] for details of the coagulation scheme) as well as grow via the nucleation and condensation of CO ice. The particle size grid in our modeling is discretized into 72 bins with a mass ratio between successive bins equal to two.

The total solid surface density in a vertical column decreases with time because particles drift radially inwards due to gas drag. Because we do not have a two-dimensional (2D) microphysical model we approximate the time evolution of the column's solid surface density analytically. At early times, the huge influx of particles from the outer disk may lead to a particle surface density given by [36]. However, at late times relevant to current observations of disks, we are in a regime where particle growth is limited by drift throughout the radial extent of the disk such that the maximum particle size present at each radial location is given by [20]:

*Equation 1*

$$a_{\text{max, drift}}(r, t_{\text{disk}}) = \frac{\Sigma_g r}{2.5 t_{\text{disk}} v_0 \rho_s}$$

where $\Sigma_g$ is the disk surface density which varies with semimajor axis, $t_{disk}$ is the current age of the system, $v_0$ approximately corresponds to the maximum drift velocity and is defined as $v_0 \equiv c_s^2/2v_k$ where $v_k$ is the local Keplerian velocity and $c_s$ is the local gas sound speed, $\rho_s$ is the internal particle density, and $r$ is the orbital distance in the disk. When particles grow larger than the size in Equation 1, we remove them from the column, reflecting the fact that they have drifted inwards.

In addition to removing particles due to drift, we also account for the influx of smaller particles due to particle drift from larger orbital distances at each time-step ($\Delta t$). After particles have grown large enough to experience drift, we calculate the flux of particles into a bin as $F_{\text{drift}} = 2\pi r v_{\text{drift}} \Sigma_g f_d$, where $v_{\text{drift}} = \Sigma_g/v_0 \rho_s a_{\text{max, drift}}$ is the particle drift velocity (in the relevant Epstein drag regime) that was used to derive Equation 1, and $f_d$ is the solid-to-gas mass ratio (including both refractories and ice) defined just exterior to the radius of interest (r+1 au) and is calculated analytically based on a treatment of particle growth following [20] (see their equation 8). We have validated that this approximation for $f_d$ is appropriate because the dust-to-gas ratio calculated using CARMA at each orbital distance closely matches the analytic expression for dust-to-gas ratio described in [20] as coagulation is the dominant mode of large particle growth in all of our models. The mass of particles that drift inwards due to drift is calculated as $F_{\text{drift}} \Delta t$. This additional mass is added to the particle bin that is smaller than the maximum particle size bin at that time. We assume that the composition of the drifting particles is equivalent to particles of that size at the given radius.

In each column, the nucleation and condensation of CO are calculated. To adapt CARMA to the diffuse outer regions of protoplanetary disks, we make several common choices for nucleation and condensation following classical nucleation theory, which has been shown to be useful in understanding experiments of the nucleation of $CO_2$ ice in the diffuse regions of Mars's atmosphere as well as the formation of water ice on Earth (e.g., [37-39], for a more detailed discussion of classical heterogeneous nucleation modeling see [40] Chapter 9). The heterogeneous nucleation process is the dominant formation pathway for CO ice formation as CO cannot undergo efficient homogeneous nucleation at the low pressures found in the outer regions of protoplanetary disks. Thus, once the initial population of CO ice-free small grains grows to sizes large enough such that nucleation is stable and ice formation can occur, CO nucleates heterogeneously. We treat a particle that has accreted CO as an ice particle unless it evaporates and releases the CO ice back to the gas phase. CARMA tracks the ice fraction of each particle throughout its evolution which enables us to self-consistently calculate mass conservation in our coupled model.

We follow [39] (see their Appendix Section A.1) to calculate the heterogeneous nucleation of CO ice on seed grains using classical nucleation theory. This prescription is the most fundamental lens of viewing heterogeneous nucleation as derived from surface energy effects. While this model is based on the formation of a liquid nucleation germ, it also serves as the basis for the heterogeneous nucleation of solids[37]. In this formulation, which includes the Kelvin effect, CO molecules diffuse over a seed particle's surface after impinging upon it until a sufficiently large number of molecules can congregate into a critical germ cluster, resulting in nucleation. We assume that the nonisothermal coefficient that accounts for the released heat of sublimation during ice growth is unity as the close contact of the forming ice and the seed particle increases the efficiency of heat dissipation for particle seeds larger than a critical cluster size (often ~$10^{-7}$ cm in our simulations) [41]. We also assume that the mean jumping distance of a CO molecule is 0.4 nm (due to a lack of experimental measurements we take the value for $CO_2$ [42]). We further assume that the density of the seed grain is the density of $CO_2$ ($\rho_{CO_2}$ = 1.5 g cm$^{-2}$), which is likely to form the initial ice mantel given the classically assumed condensation sequence in disks (e.g., [2]).

While the Kelvin effect determines whether nucleation will occur by setting the size of a critical ice cluster under a given supersaturation, a key component that determines the efficiency of heterogeneous nucleation is the geometric shape factor for ice germ formation which reduces the nucleation energy of germ formation as compared to the case of homogeneous nucleation. The shape factor appears in the exponential term of the heterogeneous nucleation rate equation and is given by[43]:

*Equation 2*

$$f(m, x) = 0.5 \left[1 + \left(\frac{1-mx}{\Phi}\right)^3 + x^3(2 - 3k + k^3) + 3mx^2(k-1)\right]$$

where x is the ratio of the radius of the ice condensation nucleus to the critical germ radius and m = cosθ is the cosine of the contact angle, θ (valued between 0º and 180º), between the ice condensate and the condensation nucleus which is a measure of their surface and interfacial energies. The use of contact angles to describe the interaction of the ice germ and the solid substrate assumes that the ice germ is a spherical cap[43,44] and assumes that there is direct vapor deposition (see [40] p. 341). We note that nucleated ice germs may have crystalline structures as may be the case for water ice (e.g., [45], p.473) such that it may be difficult to define a contact angle (although ice germs may be so small that a simple geometric description is insufficient). Furthermore, experiments demonstrate that the nucleation of water and organic crystals first occurs on specific locations (active sites) on the surface of a condensation nucleus such that specific surface features may play a dominant role in this process (e.g., [38, 46, 47]). However, the statistical model described above is still appropriate if the preferred sites for ice germ growth can be considered equal and randomly distributed[37] and this prescription remains the dominant theoretical tool to interpret laboratory experiments (e.g., [37,38,39]). Additionally, recent laboratory experiments are consistent with the predictions of classical heterogeneous nucleation theory (e.g., [48]). For example, the accretion rate of $H_2O$ and $CO_2$ ices on grains with different sizes (and thus curvatures) can be reduced by an order of magnitude for the small grain sizes relevant for protoplanetary disks[49] in comparison to much larger particles (~10,000 $\mu$m) and further depends on the composition of the ice nucleus. Furthermore, ice formation is strongly inhibited for highly curved, small grains regardless of the composition of the ice condensation nucleus. Results such as these that indicate a dependence of size and composition on the nucleation efficiency are to be expected under the picture of classical heterogeneous nucleation theory.

Given these considerations, the important parameter of the contact angle of CO ice germs on $CO_2$ ice must be determined empirically via experiments in the laboratory. Due to a lack of such experiments, we assume that the contact angle between the seed particle and the forming ice is an intermediate value of 90° such that nucleation is neither prohibited or highly efficient. This assumed value may indicate that CO is heterogeneously nucleating onto a surface where it does not preferentially form a strong bond which may be the case for CO forming on a $CO_2$ surface as indicated from laboratory experiments [50,51]. We note that when we reduce the shape factor by an order of magnitude, we initially nucleate more small particles, however, nucleation remains less efficient than condensational growth such that we ultimately derive very similar particle size distributions and the same resulting gas distributions as our nominal case. Thus, while the modeling of heterogeneous nucleation is uncertain because this process is at the limit of current theoretical knowledge, this formulation is within the range of uncertainty and can be used in this modeling due to the large separations of scale between global gaseous diffusion and hyper-local ice formation.

Once CO has heterogeneously nucleated onto a seed grain, the ice particle can grow by coagulation and/or condensation. The ice particle can also evaporate, be lofted, and/or settle to the midplane. In the scheme treated in this work, once particles coagulate, they are treated as compact grains. While the aggregate or compact nature of grains is unknown, this assumption is reasonable, particularly if condensation processes increase the compactification of particles that form as aggregates. During coagulation, when an ice particle coagulates with a seed grain, the resultant particle is treated as an icy particle.

We use a formulation of particle condensation that is applicable in the diffuse outer regions of protoplanetary disks where molecules do not diffuse along a path but instead interact collisionally, such that the change in mass of a particle with time is given by

*Equation 1*

$$\frac{dm_p}{dt} = \pi a^2 n m v_{\text{th}} \left(1 - \frac{1}{S_{\text{eq}}}\right)$$

where $m_p$ is the mass of the ice particle, $a$ is the radius of the ice particle, $n$ is the number density of condensible molecules, $m$ is the mass of a vapor molecule, and $v_{\text{th}} = (8/\pi)^{1/2} c_s$ is the thermal velocity, and $S_{\text{eq}}$ is the saturation ratio over a curved particle surface given by:

*Equation 2*

$$S_{eq} = \frac{P_{CO}}{P_{sat,eq}(a,T)}$$

The partial pressure of the CO gas, $P_{CO}$, changes with time and altitude as the system evolves.

The equilibrium condensate saturation vapor pressure depends on the size of the grain facilitating the phase change as given by

*Equation 5*

$$P_{sat,eq}(a,T) = exp\left(\frac{2\sigma v}{aRT}\right) P_{sat}(T)$$

where $P_{sat,eq}(a,T)$ is the saturation vapor pressure over a curved surface, $P_{sat}(T)$ is the saturation vapor pressure over a flat surface, $\sigma$ is the surface energy of the condensable species, $v$ is the volume per mole of the species in the condensible phase, $R$ is the universal gas constant, $a$ is the radius of the particle facilitating the phase transition, and $T$ is the temperature of the particle. The saturation vapor pressure for CO is taken from [52],

*Equation 6*

$$P_{sat}(T) = 1333.2239 \times 10^{(2.4482 - 418.44/T + 4.134 \log_{10}(T) - 0.02599T)} \text{ dynes/cm}^2$$

The growth processes in Equation 3 thus include the Kelvin effect through the supersaturation term $S_{eq}$. This formulation is adapted from the Hertz-Knudsen model for particle growth [53] with a few standard simplifying assumptions. In particular, we set the condensation and evaporation constants to be unity such that all molecules impinging on the ice particle are incorporated into the ice and evaporation is set purely by theoretical arguments and not explicitly benchmarked to laboratory data. We further assume that the ice temperature and background disk temperature are in equilibrium. For a discussion of this formulation and the simplifying assumptions we make see [54].

<u>The Kelvin effect and Particle Size Distributions</u>

The inclusion of the Kelvin effect and other surface energy effects in this work significantly alters the size distribution of ice particles from the classic view of ice formation accepted in the field. We provide a short overview of the key physical differences here and note that a more in-depth discussion will be the focus of future study.

The Kelvin effect, a well-known effect in atmospheric sciences which is important in evaluating rates of nucleation and condensation[55], including in meteorological processes on Earth[56], acts as an added barrier to ice formation on particles with high curvatures where ice coatings are unstable. This barrier is particularly stringent in the case of nucleation, which requires either the presence of large condensation nuclei or significant supersaturations of the condensable gas. While the Kelvin effect is important in regulating both nucleation and condensational growth, the most critical factor in our modeling is the barrier to nucleation. In our modeling, nucleation only occurs significantly at early times when CO gas is abundant in the ice-forming regions of the disk. For example, in our modeling, the nucleation rate decreases by 16 orders of magnitude for the same particle size when the supersaturation decreases by a factor of 5 (from S = 2000 to S = 400).

To illustrate the importance of the Kelvin effect, we modeled the same nominal disk entirely without the Kelvin effect, without the Kelvin effect in the nucleation process, and without the Kelvin effect in the process of condensational growth. The ice particle size distributions for the three cases discussed here are shown in Supplementary Figure 2. When we do not consider the Kelvin effect at all in our modeling, we uncover a broad distribution of ice particles where the ~0.1 micron grains are particularly favored sites of ice formation (Supplementary Figure 2a). This outcome is expected and can be seen in studies of the ISM (e.g., [57,58]) and previous studies of protoplanetary disks (e.g., [12,13]). In these works, the rate of ice formation on grains and as such, rates of gas-grain chemistry, is preferentially enhanced on the grains with the largest cumulative surface area (i.e., the grain size with the largest number density multiplied by its surface area). Under the assumption of an ISM-like particle size distribution, the size typically quoted is ~0.1 micron, which are particles that are large enough such that they are not readily raised to sublimation temperatures, and small enough such that they have large number densities. These small grains are readily lofted to higher altitudes in the disk where they lose their volatile material back to the gas phase such that CO gas is not as strongly depleted as in the nominal case where the Kelvin effect is considered.

When we do not include the Kelvin effect in the process of nucleation, we derive a bimodal size distribution with a mode at ~0.1 microns caused by the continuous nucleation of small grains and an extended mode at larger sizes due to condensational growth (Supplementary Figure 2b). The significant nucleation of small ice particles in this model leads to a similar outcome as the model completely devoid of the Kelvin effect where CO gas is not depleted as significantly as in our nominal case.

Finally, when we do not include the Kelvin effect in the growth process but do include the effect in nucleation, we uncover a similar outcome to our nominal case, where ice is preferentially located on the largest grains in the particle size distribution (Supplementary Figure 2c). One notable difference between this case and our nominal case is that growth occurs marginally more efficiently such that there are fewer ice grains located in the system at later times as more of them are lost to particle drift due to their enhanced size via efficient condensational growth.

It is worth noting that even with an abundant supply of background gas, the Kelvin effect causes the largest particles in a size distribution to form stable ice-coatings more efficiently than smaller grains. This effect is further exacerbated when the supply of condensable gas is diffusion-limited – as is the case in the models presented in this work. In the diffusion-limit of particle growth via condensation, large particles with the lowest barrier to ice formation preferentially form ice[55] even with their lower overall surface areas due to the smaller total number of large particles (which dominate the mass distribution of grains while small particles dominate the number density). We note that the relative inefficiency of heterogeneous nucleation as compared to condensational growth is important for atmospheric processes on Earth and is likely relevant for $CO_2$ ice formation on mars [39] and for water ice in protoplanetary disks (e.g., [85]).

Radial Diffusion and Accretion

We couple our microphysical model of non-equilibrium ice formation to a global model of gaseous diffusion. We make several simplifying assumptions in the absence of a spatially 2D microphysical model which is the subject of future work. We assume that the vertical and radial diffusion coefficients are the same at each location in the disk. The diffusion coefficient, $D$, is set by $D = \alpha c_s H$. The parameter α is a parameter of ignorance (see [59]). The sound speed and disk scale height, $H$, are defined locally at the disk midplane. We divide the disks into two hundred radial bins with an even linear spacing. We note that simulations with double the number of radial bins produce the same results.

As can be seen in Figure 1b, CO gas is only abundant in the warm upper layers of the disk which are also the regions accessible to observations [6]. We thus model the radial diffusion of CO gas in the warm layers using the 1D diffusion equation in cylindrical coordinates including radial transport due to accretion. We use a modified forward Euler scheme such that at each grid cell the mass mixing ratio of CO is set by:

*Equation 7*

$$u_i^{n+1} = g \times \left[ \left( \frac{1}{r_i} \frac{1}{\Delta r^2} \left( r_{i+1/2} D_{i+1/2} (u_{i+1}^n - u_i^n) - r_{i-1/2} D_{i-1/2} (u_i^n - u_{i-1}^n) \right) \right) dt + u_i^n + u_{\text{acc, in}} - u_{\text{acc, out}} \right]$$

where $u$ is the mass mixing ratio of CO gas at a given point in our time grid $n$ and radial distance grid $i$. The loss of CO gas due to ice formation outside of the midplane CO ice line is taken into account through the term $g = 1 - (\Delta t / \tau_{\text{vert, diff}})$, which is the fraction of the local vertical diffusion timescale that has occurred over a given time-step. The vertical diffusion timescale is calculated as $\tau_{\text{vert, diff}} = 3H^2/D$ where $H$ is defined at the disk midplane (see Supplementary Figure 3). This treatment of vertical diffusion is validated using our vertical column microphysical modeling.

While both diffusion and accretion are viscous processes that are dictated by the amount of diffusion in the system, we treat them separately because CO gas is diffusively mixed to reduce concentration gradients while also accreting following the accretion flux of the background $H_2$ gas. To calculate the terms $u_{\text{acc, in}}$ and $u_{\text{acc, out}}$, we thus determine the local accretion fluxes of $H_2$ gas and then multiply these fluxes by the local CO abundance to determine the flux in and out of the radial grid cell. Each of our $H_2$ gaseous surface density profiles are similarity solutions such that we can calculate the flux from accretion through each orbital distance grid cell using the analytic prescription described in [60] (see their Equation 21) which is appropriate for disk surface density profiles of this form. The outer disk boundary condition is set to be a zero radial concentration gradient while the inner disk boundary condition is set to be a zero disk surface density. We note that the boundary condition in the diffusion equation at the inner edge of the disk does not have an effect on our results. To accurately account for the reservoir of solid CO as a function of time we determine the dust-to-gas ratio and ice-to-dust ratio ($f_i$) as a function of time from our microphysical modeling of ice formation and linearly interpolate these values along our radial grid.

Around the midplane CO ice line, we must additionally treat the formation and drift of ice as we radially evolve the gas present in the system because there is significant cycling of particles and gas around this region [61-65]. We calculate the location of the midplane ice line as the hottest radial location in the disk where the nominal saturation ratio over a flat surface is equal to unity, i.e., the radius where $P_{CO}/P_{sat}(T) = 1$ in the midplane. Particles that drift across the midplane CO ice line quickly lose their ice mantles at a radial region just interior to the midplane ice line due to their small particle sizes [19]. Cycling occurs because some of the increased abundance of CO gas interior to the ice line will diffuse back outwards across the ice line, due to the steep radial concentration gradient. After gas diffuses across the midplane ice line, it is rapidly incorporated into solid ice due to the relatively short vertical diffusion timescale near the CO ice lines. Once more ice forms, there is again an increase in particle drift and the cycle repeats.

We model the surface density of CO ice immediately exterior to the midplane CO ice line following:

*Equation 8*

$$f_{ice} = f_d f_i + f_{accum} s$$

where $f_d$ is the analytic dust-to-gas mass ratio as a function of time described above and $f_i$ is the solid ice fraction from the microphysical modeling also described above. The term $f_{accum}$ accounts for the accumulation of CO ice due to the radial cycling of solids and gas around the ice line. It is calculated as $f_{accum}^{n+1} = f_{accum}^n + \Sigma_i/\Sigma_g - \Sigma_l/\Sigma_g$ where $\Sigma_i = (1-g)u_i^{n+1}\Sigma_g/g$ is the surface density in ice that has condensed in the last time step, $\Sigma_g$ is the surface density of the background hydrogen gas that evolves with time due to accretion, and $\Sigma_l = 2\pi r_{iceline} v_{drift} \Sigma_i^n$ is the amount of the accumulated ice surface density that has been lost to drift in the last time step. We assume that the majority of the total solid mass is located in the largest particles in the size distribution which is well supported by the results from our microphysical modeling and from previous works of drift limited particle growth [66]. We calculate the inward flux of icy pebbles at the iceline as:

*Equation 9*

$$F_{ice} = 2\pi r_{iceline} v_{drift} \Sigma_g f_{ice} .$$

We add the CO gas mass that drifts through the midplane ice line in a given timestep, $F_{ice}\Delta t$, to the radial bin just interior to the midplane CO ice line. In our models, the cycling of material around the ice line leads to an ice-to-gas ratio immediately exterior to the ice-line that is enhanced with time and a corresponding enhanced abundance of CO gas interior to the CO ice line as found in previous works [61-65].

Our simulations differ from those presented in [12, 13] as we include the accretion of gaseous CO onto the host star and allow the amount of diffusion present in the system to be a free parameter. We do not include chemical processing of CO as we find that we are able to reproduce observed levels of depletion with CO ice formation alone.

Disk Parameters

We derive the total gaseous surface densities for the disks around TW Hya and HD 163296 from previous modeling [20] using observations of dust lines. We note that the mass estimates that we use for TW Hya are consistent with the lower limit found in [5] though our favored disk mass is four times larger than the mass estimated in [67] which relies on significant modeling of the disk SED, CO, and HD observations such that the mass estimate does not represent an estimate from a single tracer such as HD observations alone. For the disk around HD 163296, we use the surface density shape derived from CO observations and a total disk. For the disk around DM Tau, which has a mass estimate from HD observations of warm molecular hydrogen which is one of the relatively direct indicators of disk mass, we normalize the surface density profile from [7] (see their Table 4 and Equation 1) such that the total disk gas mass is equal to $4.7 \times 10^{-2}$ $M_\odot$ [22] (the surface density normalization, $\Sigma_c = 0.94$ g cm$^{-2}$). For the disk around IM Lup, which does not yet have observations of a robust tracer of total disk mass, we use the gas surface density profile parameters, namely the surface density index of the gas and the gas critical radius, from [7] (see their Table 4 and Equation 1). We then choose a surface density profile normalization ($\Sigma_c = 15$ g cm$^{-2}$) such that the total disk mass falls in the range of masses described in [20] though we note that while this choice changes the derived CO mass, it has a minimal effect on the amount of depletion from the original CO abundance which is primarily controlled by diffusion.

We assume our fiducial disk temperature structures, particularly in the outer disk, are controlled by irradiation [29] (see [20] for more details). Several scale heights above the midplane, the disk is heated by high-energy photons [29]. We allow surface heating to occur in the disk two scale heights above the midplane such that the disk is isothermal within the first two scale heights $T = T_{\mathrm{mid}}(r)$. The temperature increases linearly within the third scale height of the disk until $T = 3T_{\mathrm{mid}}(r)$ and remains at this value beyond this height. This temperature structure is thus comparable to that used in [68,12,13]. To simulate the additional radiative heating that occurs due to relatively large accretion luminosities from an early phase of active disk accretion, and simultaneously improve the numerical stability of our modeling, we increase the disk temperature by a factor of two and exponentially cool the disk to rapidly reach temperatures appropriate for present day. Our results are insensitive to the timescale over which this cooling occurs (100 year and 10,000 year cooling timescales reproduces identical model results) and this process numerically serves to avoid unphysically over-shooting the amount of gas that is removed in the first timestep. We note that the exact location of surface layer heating does not impact our results as long as the first scale height remains cool enough for ice formation to occur and the overall diffusion timescale from the observed upper layers to the disk midplane is not significantly affected. This is because the level of depletion in the disk is sensitive to the saturation vapor pressure at the disk midplane, and thus the temperature at the disk midplane. The vertical density profile is in hydrostatic equilibrium calculated using these temperature profiles.

The parameters used for the disks modeled in this work are given in the Supplementary information (see Supplementary Table 1). As noted in the main text, the ages of protoplanetary disks are uncertain, and estimates vary in the literature by several million years for a given disk. The disk ages quoted in Supplementary Table 1 thus comprise a significant source of uncertainty in our modeling. We adopt these age estimates following the reasoning in [20] for the disks around HD163296 and TW Hya. IM Lup is commonly quoted to have an age of ~1 Myr (e.g., [69]). The literature around DM Tau has a broad range of age estimates from ~1 – 5 Myr ([8,70,71]). We choose to adopt an age of 1 Myr for DM Tau, as much of the literature surrounding the rapid depletion of CO gas in observed systems quotes a young age for DM Tau [8] and previous chemical models of this system were run for 1 Myr of evolution [7]. In our modeling this younger age gives rise to higher levels of diffusion that are more consistent with those inferred from observations of gas kinematics[18]. While the absolute ages of systems are uncertain, we note that their relative ages are more likely to be robust ([72]). As such, our chosen age estimates reflect that TW Hya and HD 163296 are likely significantly older than DM Tau and IM Lup.

The disk outer radii are primarily taken from [7] (see their figure 8) but are also broadly consistent with radii in [73]. For TW Hya, we adopt the outer radius from observations of $C^{18}O$ and $^{13}CO$ of 100 au as we compare to results from [6] which observe significant emission out to ~80 au. We note that increasing the outer radius of TW Hya to 215 au, in line with estimates from $^{12}CO$ emission [74], increases the CO abundance interior to ~5 au by a factor of ~2.5 due to a slightly increased rate of ice drift into the inner disk.

Comparison to Observations

For DM Tau, we compare our beam-convolved model results (red crosses) to the shape of the depletion profile from [7] (black line) as shown in Fig. 4a. We also find agreement with the total level of CO depletion in our models, a factor of 3.7, and the level of depletion reported in [22] of a factor $\lesssim 5$. For IM Lup, we compare our beam-convolved model results (purple crosses) to the shape of the depletion profiles from [7] (black line) as shown in Fig. 4c.

For TW Hya, shown in Fig. 4b, we compare our beam-convolved model results (blue crosses) to observations from [6] who report the observed CO surface density (black crosses). Our results also agree with the similar CO column density profile presented in [76]. In addition, we also compare the shape of the CO depletion profile to the shape of the gaseous CO depletion profile presented in [6] (see their figure 8). We derive a similar profile shape with a dramatic increase in CO depletion at ~20 au and a slightly reduced level of depletion exterior to this radius until the background gas disk itself becomes diffuse and truncated.

For HD 163296, shown in Fig. 4d, we compare our beam-convolved modeling results (green crosses) to observations from [31] (equivalent to the dashed black line). Our results show an overall lack of significant gas phase CO depletion though we do derive structures in the CO surface density profile. This is in agreement with the lack of disk integrated gaseous CO depletion derived in [32] using $^{13}C^{17}O$, a rare, optically thin CO isotopologue. Both [32] and [7] show a drop off in CO abundance at either ~150 au or ~70 au. Using our fiducial disk temperature profile, we find a drop off in CO abundance at ~ 80 au, although the exact location is sensitive to the assumed disk midplane temperature structure. Our modeling for this disk is in agreement with results presented in [15] which show an enhancement in CO abundance interior to the midplane ice line.

HD 163296, DM Tau, and IM Lup are likely to have optically thick CO emission from the inner edge of the thick disk out to at least ~30 au for all commonly observed isotopologues of CO such that the observed abundance or depletion factor in this region is uncertain [7]. For IM Lup in particular, Bosman et al. 2021 demonstrates that the abundance of CO interior to 30 au is unknown as the emission from $^{13}CO$ falls-off interior to this point. Their best-fit model indicates that this is likely caused by a pileup of dust grains around this region. Our model predicts such a pileup of ice grains just exterior to the midplane ice line, located around 40 au in our model (Supplementary Table 1), which is particularly pronounced in the case of IM Lup due to high levels of diffusive mixing allowing for efficient radial redistribution of volatile CO near the iceline which creates an enhancement in local ice formation. For TW Hya, gaseous CO becomes optically thick between a few and ~10 au, even for rare isotopologues like $C^{18}O$ [76]. In this region, the temperature structure is also uncertain [6], which causes the observed amount of CO in this region to be uncertain. We mark these uncertain regions in Fig. 4a-d in gray. However, while the majority of observations of CO gas in TW Hya are optically thick interior to ~20 au, we note that our results agree with the results derived from observations of an optically thin CO isotopologue in [77] from 5-21 au to within a factor of 3.

In this work, we particularly focus on the radial regions of the disks probed by $^{13}$CO and C$^{18}$O emission as these isotopologues are abundant and can serve as probes of CO column density. Emission from $^{12}$CO is likely optically thick, even in the outermost regions of disks [74] such that it is relatively insensitive to even substantial changes in column density. We also note that the disk radial scale in $^{12}$CO typically exceeds the radial scales observed in $^{13}$CO which in turn exceed the radial scales observed in C$^{18}$O (e.g., [73-75]). As shown in [75], the differences in radial scale for the different CO isotopologues may arise because the less abundant $^{13}$CO and C$^{18}$O have emission surfaces that are deeper in the disk. Thus, as the overall gas density decreases in the outer disk, $^{13}$CO and C$^{18}$O probe cooler disk regions closer to the disk midplane where CO gas is considerably depleted due to ice formation (as there is not a significant dependence on vertical mixing for the depletion of condensable gas close to the disk midplane) and their emission then quickly falls-off. Therefore, the radial scale of disks observed in $^{13}$CO and C$^{18}$O may typically be smaller than the scales observed in $^{12}$CO because the emission surfaces probe regions of the disk without significant CO abundance at shorter radii [75]. We therefore focus on the regions of the outer disk where $^{13}$CO and C$^{18}$O emit readily, as a variety of column densities may match the emission from $^{12}$CO.

Model Predictions

The primary prediction of this modeling is that ~micron sized grains are not abundantly coated in CO ice in the outer regions of evolved protoplanetary disks. Furthermore, all grains are CO ice free at heights in the disk above the level where CO gas is no longer supersaturated, which is likely below the level where UV photodesorption is active (see Fig. 1). JWST may be able to test these predictions using spatially resolved spectra of edge-on disks which may have the sensitivity to detect CO ice features that arise from small CO-coated particles and constrain their vertical distribution [24]. The detection of these features in an evolved disk would complicate the model presented here and would imply that either CO ice is abundantly present on small grains or perhaps that the CO ice forms as an aggregate with optical properties similar to small grains (e.g., [78]).

The mechanism for CO depletion presented in this work requires both vertical and horizontal diffusion within the disk. Thus, for each disk we predict a parameter $t_{\text{age}}/t_{\text{diff}}$ (see Fig. 3) where $t_{\text{age}}$ is the age of the system and $t_{\text{diff}} = r_{\text{out}}^2/\alpha c_s H$, $r_{\text{out}}$ is the outer radius of the disk, and both the sound speed and scale height are defined at $r_{\text{out}}$. Thus, for a given disk temperature structure (which sets the sound speed and scale height), an assumed disk age implies a globally averaged α mixing parameter for each disk, given in Table 1. Observations of non-thermal CO gas motions using ALMA have been used to place constraints on turbulent mixing in the upper disk layers. Our results are broadly consistent with these studies which find low upper limits for turbulent mixing in TW Hya and HD 163296 (i.e., the turbulent mixing in these systems is below observable levels) and a detectable level of non-thermal motion in DM Tau which implies a significant level of turbulent mixing [16-18]. While a direct comparison between the globally averaged α parameter derived in this work and the level of mixing in the upper disk layers may not be appropriate in a one-to-one fashion, independent observational constraints of turbulent mixing may serve to better constrain the results of the modeling presented in this work.

Another prediction of this work is that the abundance of CO gas interior to the midplane ice line is enhanced compared to the CO abundance in the outer disk. The level of enhancement will vary from system to system and primarily depends on the amount of diffusive mixing that has occurred which regulates the transport of solid CO as well as the transport and accretion of gaseous CO. We expect that future observations that constrain the CO abundance in the innermost disk regions will agree with this trend. Observations in the inner disk that point towards significant depletions comparable to those seen in the outer disk may instead point towards processes that alter solid CO into a more involatile form such as processing via grain chemistry. We further expect that systems that have undergone significant accretion of an enhanced abundance of CO gas onto the host star, like TW Hya and IM Lup, will have enhanced abundances of carbon relative to more refractory elements in the gas being accreted onto the host star. Studies of the UV spectrum of TW Hya indicate that gas-phase carbon is enhanced relative to silicon, possibly due to the accretion of carbon-rich material (e.g., [86]). Future studies of the composition of the gas accreting onto disk host stars may thus serve as an additional test of the model presented in this work (eg., [87]).

**Data Availability**
Observational CO data are published in refs. 6-7, 15, 31-32, 73-77 (see methods for more detail). Due to the large size of the data files, the full microphysical data generated by the simulations presented in this work are available from the corresponding author upon reasonable request.

**Code Availability**

The numerical models used in this work are not public. However, they are available from the corresponding author upon reasonable request.

**Acknowledgments**
 We would like to acknowledge Karin Öberg for her feedback on the surface binding properties of CO ice, Sean Andrews for his insightful discussion of the outer radii of disks as measured from CO emission, and Judit Szulagyi for insightful discussions about the temperature structures in protoplanetary disks. This work benefited from the Exoplanet Summer Program in the Other Worlds Laboratory (OWL) at the University of California, Santa Cruz, a program funded by the Heising-Simons Foundation.

**Author contributions:** D.P., P.G., and R.M.C. conceived of the project. D.P. and P.G. adapted the microphysical model of CO ice formation. D.P. coupled the radial model to the microphysical ice modeling and wrote the manuscript. D.P. and R.M.C. conceived of the concepts used in the model coupling. D.P., X.Z., R.M.C. conceived several useful tests of the finished model. All authors provided comments used in editing the manuscript.

**Funding:** D.P. acknowledges support from the Ford Foundation Dissertation Year Fellowship Program and support from NASA through the NASA Hubble Fellowship grant HST-HF2-51490.001-A awarded by the Space Telescope Science Institute, which is operated by the Association of Universities for Research in Astronomy, Inc., for NASA, under contract NAS5-26555. D.P. and R.M.C. acknowledge support from NSF CAREER grant number AST-1555385. R.M.C and X.Z. acknowledge support from NASA Interdisciplinary Consortia for Astrobiology Research (ICAR) grant 80NSSC21K0597. P. G. acknowledges support from the 51 Pegasi b Fellowship sponsored by the Heising-Simons Foundation and support from NASA through the NASA Hubble Fellowship grant HST-HF2-51456.001-A awarded by the Space Telescope Science Institute, which is operated by the Association of Universities for Research in Astronomy, Inc., for NASA, under contract NAS5-26555. X. Z. acknowledges support from the NASA Solar System Workings Grant 80NSSC19K0791 and the NASA Exoplanet Research Grant 80NSSC22K0236.

**Correspondence:** Correspondence and request for materials should be addressed to Diana Powell.

**Competing interests:** The authors declare no competing financial interests

**Supplementary Materials**

In Supplementary Figure 1 we describe the processes that shape the observed abundance of CO in protoplanetary disks. In Supplementary Figure 2 we show the size distributions of ice and ice-free particles under different ice formation schemes. In Supplementary Figure 3 we compare the vertical disk diffusion timescale with average values for the chemical processing of CO. In Fig. S4 we discuss the nucleation and growth timescales as a function of particle size. In Supplementary Table 1 we describe the model parameters used to model the four disks in our sample. In Supplementary Table 2 we describe the material properties of CO used in our microphysical ice formation modeling.

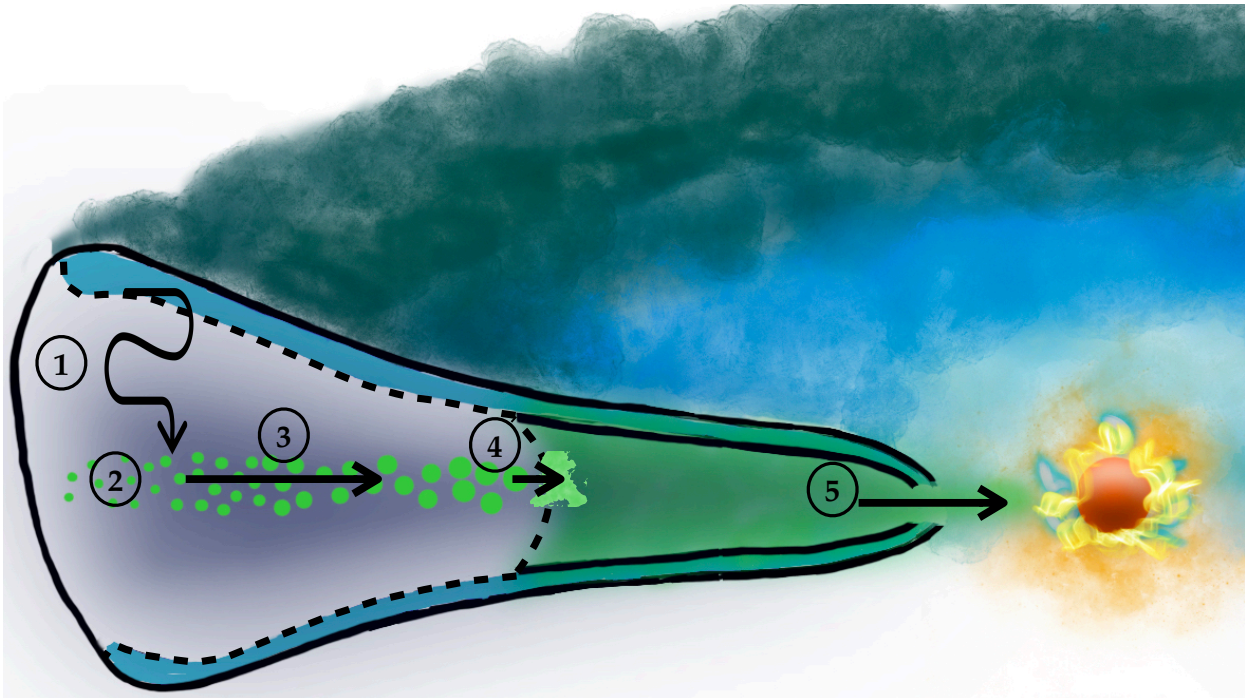

**Supplementary Figure 1.** The evolution of CO in protoplanetary disks: 1. CO gas is mixed down to the disk midplane and depleted in the warm disk layers (turquoise shaded region) above the surface ice line (black dashed line); 2. efficient ice formation sequesters CO in solids (green circles); 3. large ice coated particles drift inwards towards the host star; 4. particles lose their volatile ice content (light green shaded region) once they drift past the midplane CO ice line (black dashed line); 5. an enhanced abundance of gaseous CO (green shaded region) is accreted onto the host star. Background disk image adapted with permission from ref. [82], Arizona University Press.

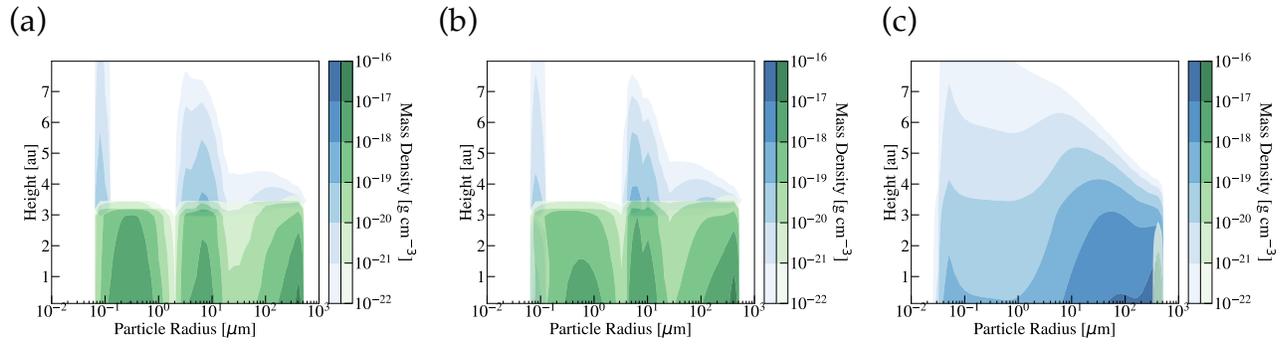

**Supplementary Figure 2**. The particle size distributions for the disk around TW Hya at 5 Myr at 30au in the case where (a) the Kelvin effect is neglected entirely, (b) the Kelvin effect is only considered for condensational growth, and (c) the Kelvin effect is only considered for nucleation. The ice-free particles are shown in blue, and the CO-ice coated particles are shown in green.

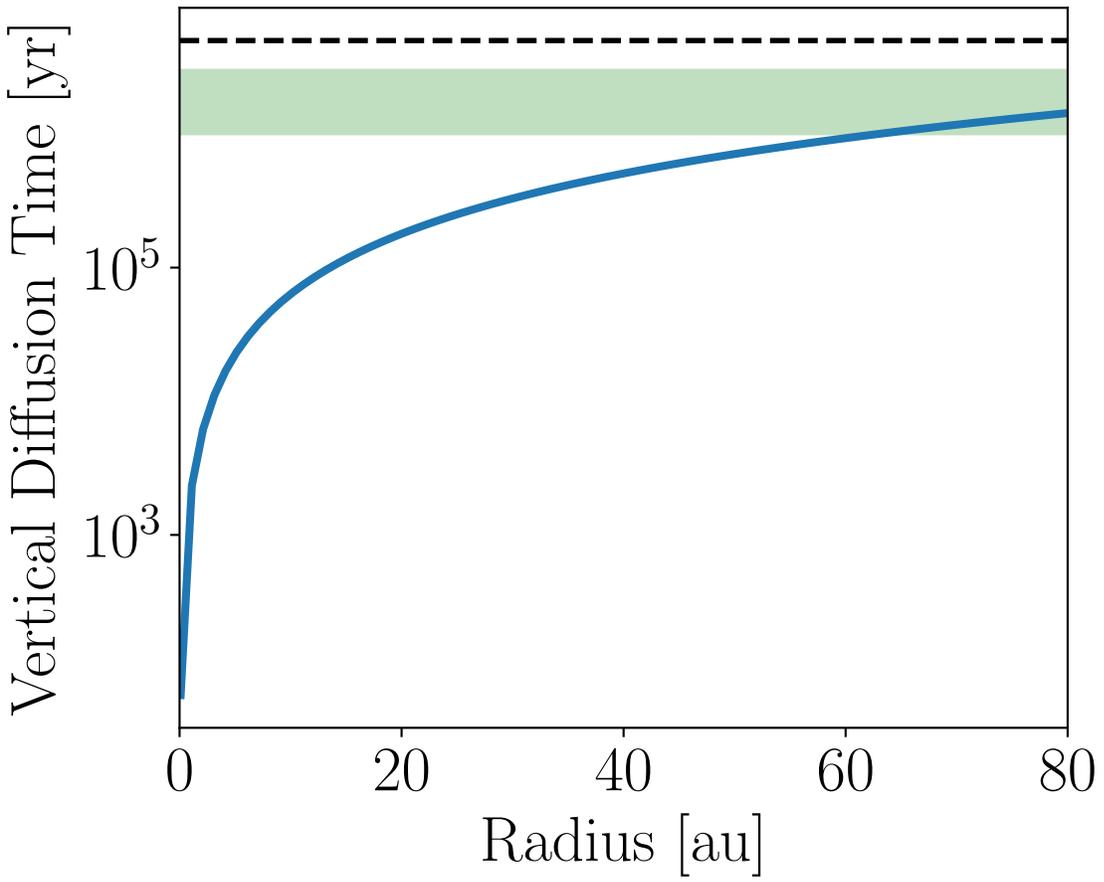

**Supplementary Figure 3**. The vertical diffusion time as a function of radius in the disk around TW Hya (blue). The diffusion timescale increases in the outer disk as the scale height increases. The vertical diffusion timescale is compared to a commonly quoted timescale for the chemical conversion of CO (e.g., ref. [10,11,14], green shaded region) is shown for comparison. The disk age is marked by the dashed black line.

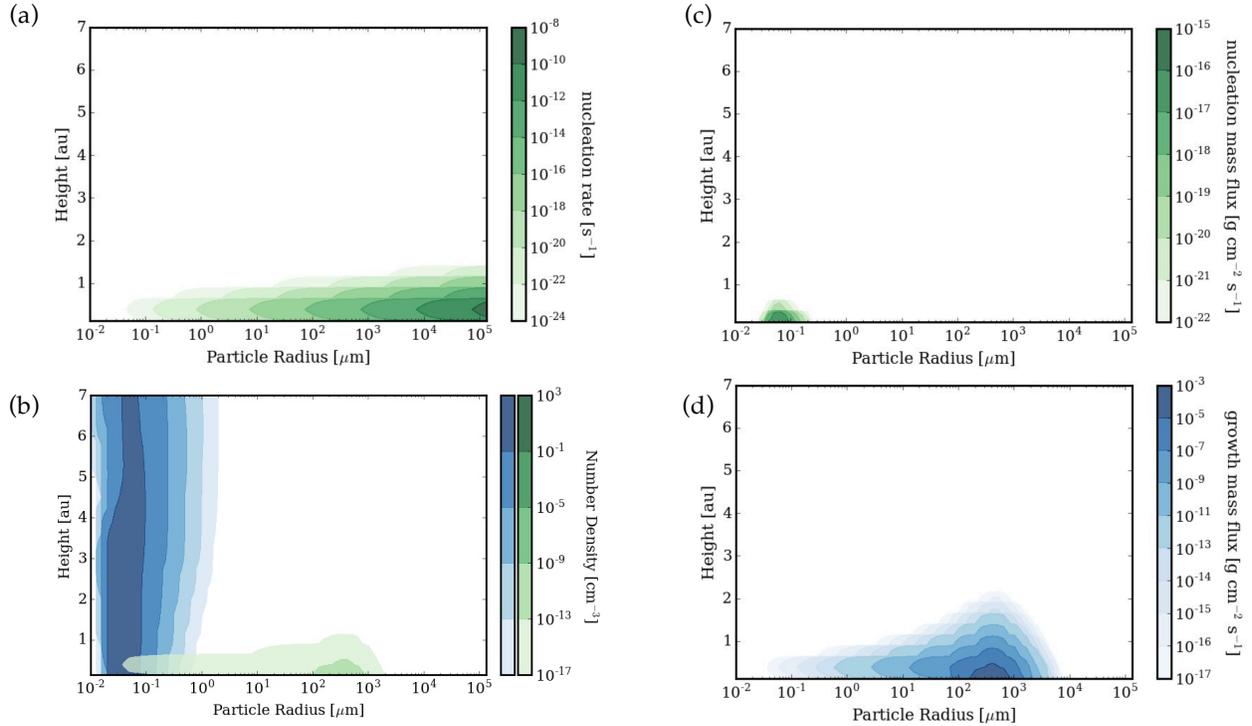

**Supplementary Figure 4**. Nucleation is much more efficient for large particles than small particles and is orders of magnitude less efficient than condensational growth. (a) The nucleation rate in units of critical germs per condensation nucleus per second in TW Hya at 30 au at a system age of ~$10^3$ years. If large particles were present at early times when CO gas is significantly supersaturated for nucleation to occur, then we would expect them to dominate the transfer of mass due to nucleation. (b) To obtain the rate at which CO gas is depleted via nucleation the nucleation rate calculated in (a) must be multiplied by the number density of ice condensation nuclei shown here for the ice-free grains. Nucleation does not occur until the particle size distribution has evolved via coagulation to a point where the median grain size approaches ~0.1 μm. The size distribution of the ice-coated grains after nucleation and subsequent condensational growth has occurred is shown for reference in green. (c) The flux of mass due to nucleation onto the existing size distribution of ice condensation nuclei. (d) The mass flux due to condensational growth far exceeds the mass flux due to nucleation, even when there are relatively few nucleated ice particles upon which condensation can occur. Note that the colorbars in plots (c) and (d) have very different scales and that the growth mass flux exceeds the nucleation mass flux by many orders of magnitude.

| Disk | $T_0$ [at 1 au] (ref. [20] see methods) | Midplane Ice Line | Disk Outer Radius (refs. [6,7] see methods) | System Age (refs. [20,8,68-69] see methods) | Stellar Mass |
|---|---|---|---|---|---|
| TW Hya | 82 K | 16 au | 100 au | 5 Myr | 0.8 $M_\odot$ (ref. [83]) |
| HD 163296 | 120 K | 44 au | 350 au | 5 Myr | 2.3 $M_\odot$ (ref. [84]) |
| DM Tau | 70 K | 19 au | 310 au | 1 Myr | 0.53 $M_\odot$ (ref. [85]) |
| IM Lup | 116 K | 43 au | 390 au | 1 Myr | 1 $M_\odot$ (ref. [86]) |

**Supplementary Table 1.**

Disk Model Parameters

| Property | Value |
|---|---|
| Surface Energy | $27.77(1 - T[K]/132.92)^{1.126}$ erg cm$^{-2}$ (ref. [87]) |
| Desorption Energy ($F_{des}$) | 1388 K (ref. [51]) |
| Condensed Density | 1.0288 g/cm$^3$ (ref. [88]) |
| Vibrational Frequency | $1.6 \times 10^{11}\sqrt{F_{des}/m_{molecular,\,CO}}$ |

**Supplementary Table 2.**
CO Material Properties

**Supplementary References:**